\documentstyle [12pt,eqsecnum,aps] {revtex}
\input epsf
\newcommand{\beq}{\begin{equation}}
\newcommand{\eeq}{\end{equation}}
\newcommand{\beqs}{\begin{eqnarray}}
\newcommand{\eeqs}{\end{eqnarray}}

\begin{document}

\draft

\baselineskip 6.0mm

\title{Ground State Entropy of Potts Antiferromagnets: \\
Bounds, Series, and Monte Carlo Measurements} 

\vspace{4mm}

\author{Robert Shrock\thanks{email: shrock@insti.physics.sunysb.edu}
\and Shan-Ho Tsai\thanks{email: tsai@insti.physics.sunysb.edu}}

\vspace{4mm}

\address{
Institute for Theoretical Physics  \\
State University of New York       \\
Stony Brook, N. Y. 11794-3840}

\maketitle

\vspace{10mm}

\begin{abstract}

   We report several results concerning $W(\Lambda,q)=\exp(S_0/k_B)$, the 
exponent of the ground state entropy of the Potts antiferromagnet on a lattice
$\Lambda$. First, we improve our previous rigorous lower bound on $W(hc,q)$ 
for the honeycomb (hc) lattice and find that it is extremely accurate; it 
agrees to the first eleven terms with the large-$q$ series for $W(hc,q)$.  
Second, we investigate the heteropolygonal Archimedean $4 \cdot 8^2$ lattice, 
derive a rigorous lower bound, on $W(4 \cdot 8^2,q)$, and calculate the 
large-$q$ series for this function to $O(y^{12})$ where $y=1/(q-1)$.  
Remarkably, these agree exactly to all thirteen terms calculated.  We also 
report Monte Carlo measurements, and find that these are very close to our 
lower bound and series.  Third, we study the effect of non-nearest-neighbor
couplings, focusing on the square lattice with next-nearest-neighbor bonds.  

\end{abstract}


\vspace{10mm}

\pagestyle{empty}
\newpage

\pagestyle{plain}
\pagenumbering{arabic}
\renewcommand{\thefootnote}{\arabic{footnote}}
\setcounter{footnote}{0}

\section{Introduction}

    Nonzero ground state disorder and associated entropy, $S_0 \ne 0$, is an
important subject in statistical mechanics; a physical realization is
provided by ice, for which $S_0 = 0.82 \pm 0.05$ cal/(K-mole), i.e., 
$S_0/k_B = 0.41 \pm 0.03$ \cite{ice}.  A particularly simple model exhibiting 
ground state entropy without the complication of frustration is the $q$-state 
Potts antiferromagnet (AF) on a lattice $\Lambda$ for sufficiently large $q$
\cite{potts}-\cite{fk}.  This subject also has a deep connection with 
graph theory in mathematics \cite{graphs}-\cite{biggsbook} since the 
zero-temperature partition function of the $q$-state Potts AF on a lattice 
$\Lambda$ satisfies $Z(\Lambda,q,T=0)_{PAF}=P(\Lambda,q)$ and hence 
$S_0/k_B = \ln W(\Lambda,q)$, where $P(G,q)$ is the chromatic polynomial, 
expressing the number of ways of coloring the vertices of a graph $G$ with 
$q$ colors such that no two adjacent vertices have the same color, and
\beq 
W(\Lambda,q) = \lim_{n \to \infty} P(\Lambda_n,q)^{1/n}
\label{w}
\eeq
where $\Lambda_n$ denotes an $n$-vertex lattice of type $\Lambda$ (with 
appropriate boundary conditions).  Given the above connection, it is 
convenient to express our bounds on the ground state entropy in terms of its 
exponent, $W(\Lambda,q)$. Recently, we studied the ground state entropy in 
antiferromagnetic Potts models on various lattices, including both Monte 
Carlo measurements and rigorous upper and lower bounds
\cite{p3afhc}-\cite{wa}. Here we continue this study. The reader is referred to
Refs. \cite{p3afhc}-\cite{ww} for further background and references. 

\section{Improved Lower Bound on $W(\lowercase{hc,q})$ for the Honeycomb 
Lattice}

   In Ref. \cite{ww}, we derived lower and upper bounds on
$W(hc,q)$ for the honeycomb (hc) lattice using a method first
applied by Biggs to obtain such bounds for the square lattice
\cite{biggs77}.  We showed that the upper and lower bounds rapidly approached
each other for large $q$ and hence became very restrictive even for 
moderately large $q$.  In particular, we obtained the lower bound, 
$W(hc,q) \ge W(hc,q)_{\ell'}$ for $q \ge 3$, where
\beq
W(hc,q)_{\ell'}= \frac{(q-1)^{3/2}}{q^{1/2}}
\label{wlbhcold}
\eeq
and found that this was very close to the actual value of $W(hc,q)$, as 
determined from Monte Carlo measurements \cite{p3afhc,w,ww},\cite{bip}. 
(We also obtained similar bounds for the triangular lattice.)  Accordingly, 
in this paper we shall focus on deriving rigorous lower bounds for several
lattices $\Lambda$, which again turn out to be very accurate not just as 
bounds but as approximations for the $W(\Lambda,q)$ functions.  In passing, we
mention that, using the same methods as in Ref. \cite{ww}, we could also derive
rigorous upper bounds for additional lattices; however, we concentrate here on
the lower bounds because of their very high accuracy as approximations for the
$W(\Lambda,q)$ functions.  We begin by 
sharpening our lower bound for the honeycomb lattice.  We picture this 
lattice as a brick lattice with the long axis of the
bricks horizontal, as in Fig. 1(a) of Ref. \cite{ww}.  We consider a
sequence of these lattices with $m$ ($n$) sites in the horizontal 
(vertical) direction, denoted $(hc)_{m \times n}$. 
In the thermodynamic limit the boundary conditions do 
not affect the bound, so for technical convenience, we take free (periodic)  
boundary conditions along the horizontal (vertical) directions.  Further, to 
maintain the bipartite nature of the lattice and avoid frustration, we take 
$n$ even.  Our bounds in Ref. \cite{ww} were obtained by taking the vectors 
of color configurations to refer to the vertical lines; here we take them to 
refer to horizontal lines $L$.  The number of allowed $q$-colorings of this 
line of vertices is ${\cal N} = P(L_m,q)=q(q-1)^{m-1}$.  One can then 
associate with two adjacent horizontal lines of sites $L$, $L'$, forming one
layer of bricks, an ${\cal N} \times {\cal N}$ dimensional symmetric coloring 
matrix $T$ with entries $T_{L,L'}=1$ or 0 if the $q$-colorings of these lines 
are or are not compatible.  Here, by compatible $q$-colorings we mean colorings
that satisfy the constraint that no two adjacent vertices have the same
color.  Then (for fixed $m,n$) $P((hc)_{m \times n},q) = 
Tr(T^n)$.  Since $T$ is a nonnegative matrix, one can apply the 
Perron-Frobenius theorem \cite{pf} to conclude that $T$ has a real positive 
maximal eigenvalue $\lambda_{max,n}(q)$.  Hence, for fixed $m$,
\beq
\lim_{n \to \infty} Tr(T^n)^{1/(mn)} \to \lambda_{max}^{1/m}
\label{trlim}
\eeq
so that
\beq
W(\Lambda,q) = \lim_{m \to \infty} \lambda_{max}^{1/m}
\label{wlim}
\eeq
Denote the column sum $\kappa_j(T) = \sum_{i=1}^{\cal N} T_{ij}$ (equal to
the row sum $\rho_j(T)=\sum_{i=1}^{\cal N}T_{ji}$ since $T^T=T$) and
$S(T) = \sum_{i,j=1}^{\cal N} T_{ij}$; note that $S(T)/{\cal N}$ is the average
row (column) sum.  Combining the bounds for a
general nonnegative ${\cal N} \times {\cal N}$ matrix $A$,
\cite{pf}
\beq
\min\{\gamma_j(A) \} \le \lambda_{max}(A) \le \max\{\gamma_j(A) \} \ ,
\quad {\rm for} \quad \gamma_j = \kappa_j \quad {\rm or} \quad \rho_j
\label{genuplowbound}
\eeq
with the ($\nu=1$ case of the ) more restrictive lower bound applicable to a
symmetric nonnegative matrix \cite{london},
\beq
\biggl [\frac{S(T^\nu)}{\cal N} \biggr ]^{1/\nu} \le \lambda_{max} \ , 
\quad {\rm for} \quad \nu=1, 2, ...
\label{lbound}
\eeq
we have
\beq
\frac{S(T)}{\cal N} \le \lambda_{max}(T) \le \max\{\kappa_j(T)\}
\label{uplowbound}
\eeq
The sum $S(T)$ is given by the chromatic polynomial for the brick layer of
length $m$, hence containing $2m$ vertices, which we calculate to be
\beq 
P((Ch)_{6,2m},q)=q(q-1)D_6(q)^{(m-1)/2}
\label{pch6}
\eeq
where the notation \cite{wa} $(Ch)_{k,2m}$ means a $2m$-vertex chain of 
$k$-sided polygons, with each adjacent pair of $k$-gons intersecting along 
one common edge, and 
\beq
D_k(q) = \sum_{s=0}^{k-2}(-1)^s {{k-1}\choose {s}} q^{k-2-s}
\label{dk}
\eeq
Taking the $m \to \infty$ limit and applying the $\nu=1$ case of 
(\ref{lbound}), we thus
obtain the lower bound $W(hc,q) \ge W(hc,q)_\ell$ for $q \ge 3$ \ \cite{bip},
where
\beq
W(hc,q)_\ell = \frac{D_6(q)^{1/2}}{q-1} = 
           \frac{(q^4-5q^3+10q^2-10q+5)^{1/2}}{q-1}
\label{wlbhc}
\eeq
For $q \ge 4$, the difference between the lower bounds (\ref{wlbhcold}) and
(\ref{wlbhc}) decreases rapidly toward zero.  The corresponding reduced $W$
functions $W_r(hc,q) = q^{-1}W(hc,q)$ have large-$q$ Taylor series which 
coincide up to order $O(q^{-4})$:
\beq
q^{-1}W(hc,q)_\ell = 1 - \frac{3}{2}q^{-1} + \frac{3}{8}q^{-2} + 
\frac{1}{16}q^{-3} + \frac{3}{128}q^{-4} + \frac{131}{256}q^{-5} + O(q^{-6})
\label{qinvwhclbtay}
\eeq
and 
\beq
q^{-1}\Bigl (W(hc,q)_\ell - W(hc,q)_{\ell'} \Bigr ) = \frac{1}{2}q^{-5} + 
O(q^{-6})
\label{qinvwhclbdif}
\eeq
Since the quantity $y=1/(q-1)$ is a natural variable for a large-$q$ expansion,
it is also useful to compare an expansion of our new lower bound with a
small-$y$ series.  For a lattice $\Lambda$, this series has the form
\cite{nagle}-\cite{kewser} 
\beq
W(\Lambda,q) = q(1-q^{-1})^{\zeta/2}\overline W(\Lambda,y)
\label{wseriesdef}
\eeq
where $\zeta$ is the lattice coordination number, and
\beq
\overline W(\Lambda,y)=1+\sum_{n=1}^\infty w_n y^n \ , \quad y = \frac{1}{q-1}
\label{wseries}
\eeq
Analogously, for the expansion of our lower bound, we define the reduced
lower bound function $\overline W(\Lambda,y)_\ell$ via 
\beq
W(\Lambda,q)_\ell = q(1-q^{-1})^{\zeta/2}\overline W(\Lambda,y)_\ell
\label{wlbseriesdef}
\eeq
 From eq. (\ref{wlbhc}), we find the very simple result 
\beq
\overline W(hc,y)_\ell = (1+y^5)^{1/2} 
\label{wbarlbhc}
\eeq
Expanding this in a Taylor series in $y$, we have 
\beq
\overline W(hc,y)_\ell = 1 + \frac{1}{2}y^5 - \frac{1}{8}y^{10} + 
\frac{1}{16}y^{15} + O(y^{20})
\label{wbarlbhctay}
\eeq
Although as a lower bound, this need not, {\it a priori}, agree with terms in
the small-$y$ Taylor series expansion of the actual function 
$\overline W(hc,y)$, we find that, remarkably, it does agree all the way up 
to $O(y^{10})$, i.e., for the first eleven terms.  The small-$y$ Taylor 
series expansion of $\overline W(hc,y)$, calculated to order $O(y^5)$ in
Ref. \cite{baker} and to $O(y^{18})$ in Ref. \cite{kewser}, is:
\beq
\overline W(hc,y)_\ell = 1 + \frac{1}{2} y^5 - \frac{1}{8} y^{10} + y^{11} + 
O(y^{12})
\label{wbaryseries}
\eeq

To show how close our lower bound is to the actual function $W(hc,q)$, we give
a comparison in Table 1 for $q=3$ through $q=10$.  We determined the values 
of $W(hc,q)$ by Monte Carlo measurements, as discussed in detail in our
previous Refs. \cite{p3afhc} and \cite{w}.  We also include a comparison with 
our slightly less stringent lower bound (\ref{wlbhcold}). Evidently, both lower
bounds are quite close to the actual measurements, even for $q$ as low as 3.
In particular, the sharper lower bound (\ref{wlbhc}) provides an extremely 
good approximation to the actual function $W(hc,q)$ for $q \ge 3$. 

\begin{table}
\caption{Comparison of lower bounds for $W(hc,q)$ and ratios of bounds to Monte
Carlo measurements for $ 3 \le q \le 10$. The estimated uncertainty from the MC
measurements for the entries in the $W(hc,q)_\ell/W(hc,q)_{MC}$ and 
$W(hc,q)_{\ell'}/W(hc,q)_{MC}$ entries is $3 \times 10^{-4}$.}
\begin{center}
\begin{tabular}{cccccc}
$q$ & $W(hc,q)_{MC}$ & $W(hc,q)_\ell$ & $\frac{W(hc,q)_\ell}{W(hc,q)_{MC}}$ 
& $W(hc,q)_{\ell'}$ & $\frac{W(hc,q)_{\ell'}}{W(hc,q)_{MC}}$ \\ \tableline
3  &  1.6600(5)   &  1.6583124  &  0.99898 & 1.6329932 & 0.98373 \\
4  &  2.6038(7)   &  2.6034166  &  0.99985 & 2.5980762 & 0.99780 \\
5  &  3.5796(10)   &  3.5794553  &  0.99996 & 3.5777088 & 0.99947 \\ 
6  &  4.5654(15)   &  4.5650849  &  0.99993 & 4.56435465 & 0.99977 \\
7  &  5.5556(17)   &  5.5552778  &  0.99994 & 5.5549206 & 0.99988 \\
8  &  6.5479(20)   &  6.5480952  &  1.00003 & 6.5479004 & 1.00000 \\
9  &  7.5424(22)   &  7.5425874  &  1.00002 & 7.5424723 & 1.00001 \\
10 &  8.5386(25)   &  8.5382220  &  0.99996 & 8.5381497 & 0.99995 \\
\end{tabular}
\end{center}
\label{wtablehc}
\end{table}

\section{$4 \cdot 8^2$ Lattice}

   Another interesting question in statistical mechanics and graph theory is
the Potts AF ground state entropy and the related asymptotic behavior of the
chromatic polynomial on heteropolygonal lattices, i.e. Archimedean lattices
\cite{arch} involving tiling of the plane by more than one regular polygon 
such that all vertices are equivalent.  In standard mathematical notation 
\cite{arch,cmo}, an Archimedean lattice is defined by the set 
$p_1 \cdot p_2 \cdot \cdot \cdot p_n$ of polygons that one traverses in a
circuit around a vertex (site) of the lattice.  For a homopolygonal lattice,
these are all the same (e.g., in this notation, the square, honeycomb, and
triangular lattices are $4^4$, $6^3$, and $3^6$).  We consider here the 
$4 \cdot 8^2$ lattice. 

\subsection{Lower Bound on $W(4 \cdot 8^2,\lowercase{q})$}

   It is convenient to represent the $4 \cdot 8^2$ lattice as a generalized
brick lattice, containing two types of bricks at each horizontal level: square
bricks and rectangular ones with (length,height) $=(3,1)$.  In the next layer 
up, a square brick is located in the center above a (3,1) brick, and a (3,1) 
brick is located with its center above a square brick, and so forth for
successive levels.  Using the same general method as before, we construct a 
coloring matrix defined between adjacent horizontal rows of bricks and obtain 
the lower bound $W(4 \cdot 8^2,q) \ge W(4 \cdot 8^2,q)_\ell$ for $q \ge
3$ \ \cite{bip}, where 
\beqs
W(4 \cdot 8^2,q)_\ell & = & \frac{\Bigl [D_4(q)D_8(q) \Bigr ]^{1/4}}{q-1} \cr
 & = & \frac{\Bigl [(q^2-3q+3)(q^6-7q^5+21q^4-35q^3+35q^2-21q+7) 
\Bigr ]^{1/4}}{q-1}
\label{wlb48}
\eeqs
Dividing by the asymptotic behavior as before \cite{ww}, this has the 
large-$q$ expansion
\beq
q^{-1}W(4 \cdot 8^2,q)_\ell = 1 - \frac{3}{2}q^{-1} + \frac{3}{8}q^{-2} 
+ \frac{5}{16}q^{-3} + \frac{51}{128}q^{-4} + O(q^{-5})
\label{wlb48tayq}
\eeq
The corresponding reduced function $\overline W(4 \cdot 8^2,y)$ is
\beqs 
\overline W(4 \cdot 8^2,y)_\ell & = & (1+y^3+y^7+y^{10})^{1/4} = 
\Bigl [ (1+y^3)(1+y^7) \Bigr ]^{1/4} \cr 
& = & (1+y)^{1/2}\Bigl [ (1-y+y^2)(1-y+y^2-y^3+y^4-y^5+y^6) \Bigr ]^{1/4} 
\label{wbar488lby}
\eeqs

\subsection{Large-$q$ Series Expansion}

    Using standard methods \cite{nagle}, we have calculated the large $q$ 
expansion, i.e., the small-$y$ expansion, of the reduced function 
$\overline W(4 \cdot 8^2,y)$.  We find 
\beq
\overline W(4 \cdot 8^2,y) = 1 + \frac{1}{4}y^3 - \frac{3}{2^5}y^6 
+ \frac{1}{4}y^7 + \frac{7}{2^7}y^9 + \frac{1}{2^4}y^{10} 
- \frac{77}{2^{11}}y^{12} + O(y^{13}) 
\label{wbar488series}
\eeq
Next, in order to assess the accuracy for large $q$ of our lower bound, we have
expanded the corresponding reduced lower bound function 
$\overline W(4 \cdot 8^2,y)_\ell$, eq. (\ref{wbar488lby}) in a series in 
$y$.  Remarkably, the small-$y$ series expansion of our lower bound agrees to 
$O(y^{12})$, i.e. to all thirteen terms that we have calculated, with the 
small $y$ expansion of the true $\overline W(4 \cdot 8^2,y)$ function in
(\ref{wbar488series}) !  This shows that for even moderately large $q$, our 
lower bound $W(4 \cdot 8^2,q)_\ell$ actually provides an extremely accurate 
approximation to the true function $W(4 \cdot 8^2,q)$.  Of course, it is
straightforward to calculate the series (\ref{wbar488series}) to higher order
in $y$, and at $O(y^{13})$ or above the series for $\overline W(4 \cdot 8^2,y)$
and $\overline W(4 \cdot 8^2,y)_\ell$ could differ; nevertheless, the 
agreement to $O(y^{12})$ is already quite striking. 

\subsection{Monte Carlo Measurements}

   To get further information on $W(4 \cdot 8^2,q)$, we have carried out Monte
Carlo (MC) measurements of the Potts AF ground state entropy $S_0$ and hence
obtained $W = \exp(S_0/k_B)$, for $3 \le q \le 10$.  Our methods are similar 
to those used in our previous works \cite{p3afhc}-\cite{w}.  We present our
results in Table 2, including comparisons with our rigorous lower bound and our
series calculation to $O(y^{12})$ (to which order they coincide).  Evidently,
the agreement is extremely good, even for $q$ as low as 3. 

\begin{table}
\caption{Exponential of ground state entropy, from Monte Carlo measurements,
and comparison with lower bound, for the Potts antiferromagnet with $4 \le q
\le 10$ on the $4 \cdot 8^2$ lattice.  The estimated uncertainty from the MC
measurements for the entries in the 
$W(4 \cdot 8^2, q)_\ell/W(4 \cdot 8^2,q)_{MC}$ column is $3 \times 10^{-4}$.}
\begin{center}
\begin{tabular}{ccc}
$q$ & $W(4 \cdot 8^2,q)_{MC}$ &
$\frac{W(4 \cdot 8^2, q)_\ell}{W(4 \cdot 8^2,q)_{MC}}$ \\ \tableline
3  &  1.68575(60)    &  0.99959  \\
4  &  2.62226(75)    &  0.99994  \\
5  &  3.5918(10)     &  0.99995  \\
6  &  4.5737(14)     &  0.99996  \\
7  &  5.5618(16)     &  0.99992  \\
8  &  6.5530(20)     &  0.99995  \\
9  &  7.5466(23)     &  0.99995  \\
10 &  8.5413(26)     &  0.99998  \\
\end{tabular}
\end{center}
\label{wtable48}
\end{table}

\section{Square Lattice with Diagonal Next-Nearest-Neighbor Couplings}

   An interesting question concerns the effect of non-nearest-neighbor 
interactions on the ground-state entropy of the Potts antiferromagnet and the
corresponding $W(\Lambda,q)$ function. We explore this question here by 
considering the Potts AF on a square lattice with next-nearest-neighbor 
couplings along both diagonals, defined by the Hamiltonian 
\beq {\cal H} = -\sum_{n,n'} J_{nn'} \delta_{\sigma_n, \sigma_{n'}}
\label{pnnnham}
\eeq
where $J_{n,n'}=J < 0$ if the sites $n$ and $n'$ are nearest neighbors along a
row or column, and $J_{n,n'}=J_d < 0$ if $n$ and $n'$ are diagonally opposite 
sites on a given square of the lattice (with each such pair counted only once 
in the sum (\ref{pnnnham})).  We further define $K = \beta J$ and $K_d = 
\beta J_d$ and consider the $T \to 0$ limit, so that the color on each site $n$
is required to be different from the colors of each of its nearest neighbors 
on the rows and columns of the lattice, and also each of its next-nearest
neighbors along the diagonals of the lattice.  Then $Z((sq)_d,q,T=0)_{PAF} = 
P((sq)_d,q)$, where $(sq)_d$ denotes the nonplanar graph formed from a square 
lattice by adjoining bonds connecting diagonally opposite sites on each square
\cite{unionjack}.  Note that the $(sq)_d$ lattice has coordination number
$\zeta=8$ and girth $\gamma=3$, where ``girth'' $\gamma$ of a graph $G$ 
denotes the length of the shortest circuit on $G$.  We find that the chromatic
number of this lattice (for free boundary conditions or periodic boundary
conditions that preserve the bipartite structure of the underlying square
lattice before inclusion of diagonal bonds) 
\beq
\chi((sq)_d)=4
\label{chisqd}
\eeq
($\chi(G)$ is defined in footnote \cite{bip}.)  These are to be compared with 
$\zeta = \gamma = 4$ and $\chi(sq)=2$ for the square lattice.  From our 
earlier discussion, one thus has $W((sq)_d,q) \le W(sq,q)$. Indeed, we find 
that there is an especially strong difference with respect to the square
lattice for the case $q=\chi((sq)_d)=4$.  Here, we calculate that for a 
$(sq)_d$ lattice  $\Lambda_{L_1 \times L_2}$ of length $L_1$ in the $x$ 
direction and $L_2$ in the $y$ direction,
\beq
P(\Lambda_{L_1 \times L_2},4) \sim 2^{L_1 + L_2}
\label{psqdasymp}
\eeq
in the thermodynamic limit.  Although this diverges, it does not diverge
rapidly enough to yield a finite entropy.  With the ordering of limits 
\beq
W(\Lambda,q_s) \equiv \lim_{n \to \infty} \lim_{q \to q_s} P(G,q)^{1/n}
\label{wdefnq}
\eeq
as discussed in Ref. \cite{w}, we thus obtain the exact result
\beq
W((sq)_d,q=4) = 1 \qquad i.e., \qquad S_0((sq)_d,q=4)=0
\label{wsqdq4}
\eeq
This may be compared with the value for the usual square lattice, for which 
$W(sq,4)=2.3370(7)$ \cite{w}, corresponding to the nonzero ground state
entropy $S_0(sq,4)=0.8489(3)$.  (Indeed, even without any detailed calculation,
one knows that $S_0(sq,4)$ is nonzero from the elementary rigorous lower 
bound on a bipartite lattice, $S_0(\Lambda_{bip.},q) > (1/2)\ln(q-1)$.) 

\begin{table}
\caption{Exponential of ground state entropy, from Monte Carlo measurements,
and comparison with lower bound, for the Potts antiferromagnet with $5 \le q
\le 10$ on the square lattice with diagonal next-nearest-neighbor couplings.
The estimated uncertainty from the MC measurements in the entries for 
$W(sq_d,q)_\ell/W(sq_d,q)_{MC}$ is $1 \times 10^{-3}$.}
\begin{center}
\begin{tabular}{cccc}
$q$ & $W(sq_d,q)_{MC}$ & $\frac{W(sq_d,q)_\ell}{W(sq_d,q)_{MC}}$ & 
$W(sq,q)_{MC}$ \\ \tableline
5  &  1.5781(16)  &  0.95051 & 3.2510(10)   \\
6  &  2.4460(24)  &  0.98119 & 4.2003(12)   \\
7  &  3.3660(33)  &  0.99029 & 5.1669(15)   \\
8  &  4.3093(43)  &  0.99453 & 6.1431(20)   \\
9  &  5.2680(53)  &  0.99658 & 7.1254(22)   \\
10 &  6.2363(62)  &  0.99774 & 8.1122(25)   \\
\end{tabular}
\end{center}
\label{wtablesqd}
\end{table}

    Using the same type of coloring matrix method as above, we obtain the
rigorous lower bound $W((sq)_d,q) \ge W((sq)_d,q)_\ell$, where 
\beq
W((sq)_d,q)_\ell = \frac{(q-2)(q-3)}{q-1} \qquad {\rm for} \quad q \ge 4
\label{sqdlb}
\eeq
For large $q$, we have the Taylor series expansion 
\beq
q^{-1}W(sq_d,q)_{\ell} = 1 - 4q^{-1} + 2( q^{-2} + q^{-3} + q^{-4} + 
O(q^{-5}))
\label{ser}
\eeq
 From eq. (\ref{wlbseriesdef}) we have
\beq
\overline W(sq_d,y)_\ell = (1-y)(1-2y)(1+y)^3
\label{wsqdylb}
\eeq

The MC measurements are given in Table 1.  For comparison, we include also our
MC measurements of $W(sq,q)$ \cite{w}.  The lower bound (\ref{sqdlb}) steadily
approaches the measured values $W((sq)_d,q)$ as $q$ increases and is quite 
close to them for $q$ greater than about 6. The inequality $W((sq)_d,q) \le
W(sq,q)$ is also evident.

This research was supported in part by the NSF grant PHY-93-09888.

\vfill
\eject

\begin{thebibliography}{99}
\bibitem{bip}{We comment on the case $q=2$ here.  A bipartite lattice
$\Lambda_{bip.}$ has chromatic number $\chi(\Lambda_{bip.})=2$, where $\chi(G)$
is defined as the minimum integer value of $q$ that allows a $q$-coloring of
$G$ such that any two adjacent vertices have different colors.   In the 
present work, the honeycomb and $4 \cdot 8^2$ lattices are bipartite. 
In general, there are two ways to color a bipartite lattice with $q=2$ 
colors (provided that the boundary conditions maintain the bipartite 
property).  For $q=2$, one does not need any bounds since one has an exact
result: in the subscript notation of Ref. \cite{w}, 
$W(\Lambda_{bip.},2)_{D_{nq}} = 
\lim_{n \to \infty} \lim_{q \to 2} P((\Lambda_{bip.})_n,q)^{1/n}=1$; 
equivalently stated, the $q=2$ Potts AF (Ising model) on $\Lambda_{bip.}$ 
has no ground state entropy. }
\bibitem{unionjack}{This nonplanar graph differs from the (planar) union 
jack lattice in that the diagonals do not form a further vertex in the middle
of each square.}
\end{thebibliography}
\end{document}